\documentclass[12pt]{article}

\def\be{\begin{equation}}
\def\ee{\end{equation}}
\def\ba{\begin{array}}
\def\ea{\end{array}}
\def\de{\partial}

\def\calG{{\cal G}}
\def\calH{{\cal H}}
\def\calF{{\cal F}}
\def\ds{{\displaystyle}}

\begin{document}

\title{Generalized regularly discontinuous solutions of the Einstein equations}

\author{Gianluca Gemelli}
\date{{\small L.S. B. Pascal, V. P. Nenni 48, Pomezia (Roma), Italy. \\
e-mail: gianluca.gemelli@poste.it}}

\maketitle

\begin{abstract}
The physical consistency of the match of piecewise-$C^0$ metrics is discussed. The mathematical theory of gravitational discontinuity hypersurfaces is generalized to cover the
match of regularly discontinuous metrics. The mean-value differential geometry framework on a
hypersurface is introduced, and corresponding compatibility conditions are deduced. Examples of generalized boundary layers, gravitational shock waves and thin shells are studied.
\end{abstract}

Submitted to Int. J. Theor. Phys.


\section{Introduction}

Is it possible to define weak solutions of the Einstein equations of class
piecewise-$C^0$, i.e. to generalize the compatibility conditions which replace the field equations on a singular hypersurface to the case when the metric is regularly discontinuous? 

To reach this goal would probably mean to define the most general class of regularly discontinuous weak solutions of the Einstein equations. It seems that this problem was never studied before in the literature. But, before we proceed, we need to discuss whether we are talking of something  mathematically and physically consistent or not. 

A fundamental concept of Riemannian geometry is that at any point of a
submanifold there are coordinate choices for which the metric reduces to the
Minkowski flat metric. Clearly, if this choice is made
on both sides of the discontinuity surface, any "jump" in the metric disappears. Thus, the 
metric discontinuity appears as a 
coordinate dependent concept, which is neither geometrically (nor 
physically) acceptable in the context of General Relativity. 

But we also have to consider that regularity of the global coordinates plays an important role in our approach, which is that of \cite{iogrg2002} and of the literature cited therein. In particular, since here the spacetime is only $C^0$, we are led to considering ($C^0$, piecewise $C^1$)
coordinate transformations. If the metric is discontinuous in some
globally $C^0$ chart, it is in general impossible to obtain the vanishing of the metric jump on both sides of a hypersurface with a $C^0$ coordinate transformation (see section 2). Moeover in the following we are led in a natural way to considering $C^1$ coordinate transformations; the metric discontinuity is a tensor with respect to such coordinate changes!

In other words the jump of the metric has a precise mathematical meaning, if we consider it in connection with global regular coordinates. 

In a well consolidated procedure, the assumption of continuity for the metric across a gravitational interface is usually taken for granted; however it follows from the limiting process of the thin sandwich modelization, in consequence of the hypothesis that the external derivatives of the metric are bounded \cite{obr}. Yet in this paper we are going to see that,
even removing the assumption of continuity, it is still possible to define a generalized inner geometry of the discontinuity hypersurface; one thus can consistently find a corresponding generalized set of compatibility conditions, which obviously reduces to the usual ones when the continuity hypothesis is restored.  

Yet, which are the physical motivations to move to such generalization? 
Actually gravitational shock waves and thin shells are usually defined by the presence of singular curvature with a ``delta" component concentrated on a hypersurface, situation which is well cast within the classic $C^0$ piecewise-$C^1$ match of metrics \cite{iogrg2002,barrabeshoganbook}. 

We were originally led to consider solutions of class piecewise-$C^0$, as possible generalizations of shock waves and thin shells, by the sake of mathematical completeness, with the idea that phisical interpretation would follow. 
We actually found a reacher framework than the usual one, with some interesting new features (and even some rather undesiderable ones), which we display in this paper.

There are two main theories in the literature
for solutions of class $C^0$ piecewise-$C^1$, i.e.
that in terms of the second fundamental form (heuristic theory, see e.g. \cite{israel66,barra91})
and that in terms of the curvature tensor-distribution (axiomatic theory, see e.g. \cite{lich94,iogrg2002}); such are equivalent
through appropiate extensions (for a 
self-contained overview see e.g. \cite{iogrg2002}). 

The axiomatic theory appears to be inappropriate to the study of generalized solutions,
since the theory of distributions is basically linear. Even if we could in principle replace the discontinuous metric with its associated distribution $g^D$, then it would be impossible to define, for example, replacements for the Christoffel symbols, since this would involve product of distributions, which, as it is generally believed, is impossible to define. In fact it was proved by Schwarz \cite{schwarz54} that, under reasonable hypothesis, there can be no definition of commutative and associative operation on distributions which reduce to ordinary multiplication on integrable distributions (say on regular functions); thus in a word it is impossible to define product of distributions. 

Or is it? Colombeau \cite{colo83,colo84,colo92} developed a theory which apparently contradicts Schwarz's result. He introduced a very broad space of generalized functions, which extends the usual space of distributions, a subspace of which corresponds, in a certain sense (the correspondance is not 1 to 1), to usual distributions. Colombeau's formalism 
permits multiplication of generalized functions; but the
contradiction with the impossibility theorem is only apparent, 
in fact Schwarz's hypothesis are violated, since the operation does not coincide with ordinary multiplication on regular functions nor with multiplication of a regular function times a distribution (although it does at least for $C^\infty$ functions). 

Such theory, however, does not fit in a natural way in general relativity, since it is impossible to define covariantly invariant geometrical objects; in fact Colombeau's space is not invariant for smooth coordinate transformations, unless they are linear. Such difficulty, however, seems to have been overcome in subsequent adjustments of the theory, with the introduction of a richer mathematical framework \cite{colo94,vickers98}, so that the generalized functions current apparatus can be used in general relativity, 
and indeed it has been applied at least to the calculation of singular curvatures of the spacetimes of Kerr \cite{balasin97}, Reissner-Nordstrom \cite{stein97},
and so-called cosmic-string spacetime \cite{clarke96}.  
In such literature Colombeau's theory is adapted to the handling of curvature when the metric has a singularity in the sense of functions, i.e. the ordinary curvature would explode, at a singular event-point or at a singular worldline. There seems to be no particular reason to forbid Colombeau's method also for defining the match of piecewise-$C^0$ regularly discontinuous metrics at a singular hypersurface; however, as far as the author is aware, no attempt has been made yet to use it in this framework.

The direct method we will introduce in the following sections, however, is so conceptually simple that we prefer not to experiment with Colombeau's generalized functions, which would instead mean introducing a far more complicated and unfamiliar mathematical apparatus.  

In this paper in fact we propose a new, generalized theory for regularly discontinuous solutions, covering also the match of piecewise-$C^0$ metrics. Our theory is heuristic, as it is constructed in a way similar to
the heuristic theory of $C^0$, piecewise-$C^1$ solutions originated from the studies of Israel, but we completely avoid the traditional or projectional Gauss-Codazzi framework (which either does not include the lightlike case \cite{israel66,barra91}, or needs a special adaptation for it \cite{iojgp2002,iogrg2002}) and introduce what
we called ``mean-value differential geometry" framework, instead (see section 3). This is conceptually very simple, and permits to construct in a natural way a generalized theory, where the main role (which used to be that of the jump of the secund fundamental form) is here played by the jump of the Christoffel symbols.
The theory is an extension of the theory of gravitational discontinuity hypersurfaces we have
studied in \cite{iogrg2002}, to which it reduces when the metric is $C^0$. Even if we should restrict to $C^0$ solution, by adding the traditional assumption
of continuity for the metric, our theory would undoubtedly have at least the good qualities of not needing the timelike and the lightlike case to be distinguished (different from usual heuristic theory), and of just requiring $C^0$ continuity for the coordinates (different from the axiomatic theory). Moreover, it is completely cast in the framework of general coordinates of the ambient (glued) spacetime, with no use of parametric equations of the hypersurface, nor of inner coordinates and holonomic 3-basis, which could be considered a good quality in some applications as well. 

Piecewise-$C^0$ weak solutions of the Einstein equations, 
as far as the author is aware, have never been
considered previously in the literature. They generalize the corresponding $C^0$ solutions, as examples in this paper show; however there is more. Apparently in fact the theory allows the propagation of free gravitational discontinuity at lower speed than the speed of light (section 8); or rather, we still have no general proof that the absence of stress energy concentrated on $\Sigma$ should, in the timelike case, necessarily imply the degeneracy of a generalized solution to a boundary layer, although it does at least for a wide class of spherical matchs (see section 6).
Moreover, non-simmetric stress-energy is allowed on the hypersurface (section 9), like e.g. in Einstein-Cartan dynamics. This possible link to classical unification theories is surprising, since in our framework we have nothing similar to Einstein-Cartan torsion.
We therefore see a lot of space for future investigation.

\section{Discontinuous metrics}

Let us suppose $V_4$  
an oriented differentiable manifold of dimension 4, 
of class ($C^0$, piecewise $C^2$), 
provided with a strictly hyperbolic metric of signature --+++ and 
class piecewise-$C^0$. 
Let $\Omega \subset V_4$ be an open connected subset with compact 
closure. Let units be chosen in order to have the speed of light in empty space 
$c\equiv 1$. Greek indices run from 0 to 3.

Let $\Sigma \subset \Omega$ be a regular hypersuperface of equation 
$f(x )=0$; let $\Omega^+$ and $\Omega^-$ denote the subdomains 
distinguished by the sign of $f$. 
We suppose the metric and its first and second
partial derivatives to be
regularly discontinuous on $\Sigma$ in all charts of class $C^0(\Omega)$.
Let $f\in C^0(\Omega )\cap C^2(\Omega\backslash\Sigma )$, and let second 
and third derivatives of $f$ be regularly discontinuous on 
$\Sigma$. Finally, let $\ell_\alpha\equiv\de_\alpha f$ denote the gradient of 
$f$. 

Let the metric be a solution of the ordinary 
Einstein equations on each of
the two domains $\Omega^+$ and $\Omega^-$. 
In this situation $\Sigma$ is the interface between two general
relativistic spacetimes and it is called a (generalized) gravitational 
discontinuity hypersurface.

In the following we will develope a theory to justify the introduction of suitable generalized compatibility conditions to replace the Einstein equations on $\Sigma$ (section 5); if such conditions are satisfied the match across the generalized gravitational hypersurface $\Sigma$ will be called a {\it generalized regularly discontinuous solution of the Einstein equations}.  

Now let us briefly recall some basics notions on regularly discontinuous fields, which
we will use as tools. In any case, for notation and terminology we refer to \cite{iogrg2002}.  

A field $\varphi$ is said to be regularly discontinuous on $\Sigma$ if
its restrictions to the two subdomains $\Omega^{+}$ and $\Omega^{-}$ 
both have a finite limit for $f\longrightarrow 0$; we denote such limits by 
$\varphi^+$ and $\varphi^-$,
respectively.  

In this case the
{\it jump} $[\varphi]$ across $\Sigma$ and its {\it arithmetic mean value} $\overline{\varphi}$ are well defined on the hypersurface:

\be
\ba{lcl}
[\varphi]&=&\varphi^+-\varphi^-\cr
\overline{\varphi}&=&(1/2)(\varphi^++\varphi^-)
\ea
\ee 
If $\varphi$ is continuous across $\Sigma$, we obviously have:
$[\varphi]=0$, $\overline{\varphi}=\varphi$. 
We also have the converse formulae:

\be
\ba{lcl}
\varphi^+&=&\overline{\varphi}+(1/2)[\varphi]\cr
\varphi^-&=&\overline{\varphi}-(1/2)[\varphi].
\ea
\label{conversejumps}
\ee
As for the product of two functions $\varphi$ and $\psi$, we have: 

\be
\ba{lcl}
[\varphi \psi]&=&[\varphi]\overline{\psi}+\overline{\varphi}[\psi]\cr
\overline{\varphi\psi}&=&\overline{\varphi}\overline{\psi} +(1/4)[\varphi][\psi]
\ea
\label{product}
\ee
If a field $\varphi$ is regularly discontinuous on $\Sigma$,
its jump $[\varphi]$ is sometimes called its {\it discontinuity} of order 0.

The jump of a regularly discontinuous 
function has support on $\Sigma$, but
in general, the partial derivative of the jump is well defined as the jump of the derivative of the function
(see \cite{cat78,iojgp96}). In particular, the derivative of the jump of
a continuous field is not null, unless the field is also $C^1$. 

Similarly, we define the partial derivative of the mean value
as the mean value of the partial derivative.
We can also use regular prolongations to extend, in a sense, the definition of $\overline{\varphi}$ and $[\varphi]$ to the whole domain $\Omega$. Thus they can be regarded as regular and derivable fields in $\Omega$, but their values (and those of their derivatives) are well defined only on $\Sigma$, while in $\Omega\backslash\Sigma$ they depend on the choice of the prolongation. For details on the method of regular prolongations see e.g. \cite{cat78,iojgp96}.

We moreover define the covariant derivative of a field with support on $\Sigma$
by means of the mean value $\overline{\Gamma}_{\beta\rho}{}^\sigma$ 
of the Christoffel symbols.
For the jump of a regularly discontinuous vector, for example, with 
this definition one has that the jump of the covariant derivative is different than the
covariant derivative of the jump.
Thus, by definition, we have:

\be
\nabla_\alpha[V^\beta]= \de_\alpha[ V^\beta] + \overline{\Gamma}_{\alpha\sigma}{}^\beta[V^\sigma]
\label{defcdweak}
\ee
and in consequence of (\ref{product}):
\begin{eqnarray}
\nabla_\alpha[V^\beta]
=[\nabla_\alpha V^\beta] - [\Gamma_{\alpha\sigma}{}^\beta]\overline{V}^\sigma \ ,
\label{covariantj}
\end{eqnarray}
and similarly for the jump of any regularly discontinuous tensor.

Since the spacetime is only $C^0$, we are led to considering ($C^0$, piecewise $C^1$)
coordinate transformations, with regularly discontinuous first derivatives; the metric discontinuity $[g_{\alpha\beta}]$ is not a tensor with respect to such changes of coordinates. In fact we have:

\be
[g_{\alpha\beta}]=[g_{\alpha'\beta'}]\overline{\frac{dx^{\alpha'}}{dx^\alpha}} \cdot \overline{\frac{dx^{\beta'}}{dx^\beta}} +q_{\alpha\beta'} \Biggl[\frac{dx^{\beta'}}{dx^\beta}\Biggr] +q_{\alpha'\beta} \Biggl[\frac{dx^{\alpha'}}{dx^\alpha}\Biggr]
\label{giggione}
\ee
where:
\be
q_{\alpha'\beta} = \frac{1}{8}[g_{\alpha'\beta'}]\Biggl[\frac{dx^{\beta'}}{dx^\beta}\Biggr]+\bar{g}_{\alpha'\beta'}\overline{\frac{dx^{\beta'}}{dx^\beta}} 
\ee
We therefore can simulate all ($C^0$, piecewise $C^1$) coordinate changes by combining $C^1$ changes with {\it metric gauge} changes:

\be
[g_{\alpha\beta}] \longleftrightarrow [g_{\alpha\beta}] +q_{\alpha\beta'} \Biggl[\frac{dx^{\beta'}}{dx^\beta}\Biggr] +q_{\alpha'\beta} \Biggl[\frac{dx^{\alpha'}}{dx^\alpha}\Biggr]
\ee
which generalize usual gravitational gauge changes of the theory of ($C^0$, piecewise $C^1$) solutions \cite{iogrg2002}.

Is it always possible to make $[g_{\alpha\beta}]$ vanish with an appropriate $C^0$ transformation? Clearly the answer is negative. In fact it suffices to consider the case when $[g_{\alpha\beta}]$ and $\bar{g}_{\alpha\beta}$ are both definite positive in a given chart to see that the equation obtained from 
(\ref{giggione}) by replacing the first hand side with 0 has no solution for $[\de x^{\alpha'}/\de x^\alpha]$
and $\overline{\de x^{\alpha'}/\de x^\alpha}$. Thus the set of effective generalized gravitational discontinuity hypersurfaces is non empty. 

Moreover it will occur in many applications to have $\ell_\alpha\in C^0$. Therefore it will be often desiderable to work in the framework of ($C^1$, piecewise $C^2$) coordinate transformations, which preserve such condition. The metric discontinuity is a tensor with respect to such changes
of coordinates, but the jump of the Christoffel symbols, which appear to play a main role in the following, is not; we have in fact:

\be
[\Gamma_{\alpha\beta}{}^\sigma]=[\Gamma_{\alpha'\beta'}{}^{\sigma'}]\frac{\de x^{\alpha'}}{\de  x^\alpha}\frac{\de x^{\beta'}}{\de x^\beta}\frac{\de x^\sigma}{\de x^{\sigma'}}+\biggl[\frac{\de^2x^{\sigma'}}{\de x^\alpha\de x^\beta}\biggr]\frac{\de x^\sigma}{\de x^{\sigma'}}
\ee 
If the coordinates are $C^0$ and so is the form $\ell_\alpha$ we can write:

\be
\biggl[\frac{\de^2x^{\sigma'}}{\de x^\alpha\de x^\beta}\biggr]=\ell_\alpha\ell_\beta\de^2 x^{\sigma '}
\ee
where $\de^2$ denotes the weak discontinuity of order 2 (see  e.g. \cite{cat78,iojgp96}). Thus on $\Sigma$ we can generate all ($C^1$, piecewise $C^2$) transformations for $[\Gamma]$ combining $C^2$ transformations (with respect to which $\Gamma$ is a tensor) and {\it Christoffel gauges transformations}, i.e. of
the kind:

\be
[\Gamma_{\alpha\beta}{}^\sigma] \leftrightarrow [\Gamma_{\alpha\beta}{}^\sigma]+\ell_\alpha \ell_\beta Q^\sigma
\label{gauge}
\ee 
with some analogy with the case of $C^0$ metrics (where the main role is played by the first order metric discontinuity $\de g$, see \cite{iogrg2002} section 3).

In any case neither the mean value of the metric $\overline{g}$ or its jump $[g]$ now have null covariant derivatives. 
Consider in fact the identity $\nabla_\alpha g_{\beta\rho}=0$ in the domain $\Omega^+$; from the limit
$f\longrightarrow 0^+$, on $\Sigma$ we have:

\be
\de_\alpha g_{\beta\rho}^+ -(\Gamma_{\alpha\beta}{}^\nu)^+g_{\nu\rho}^+-(\Gamma_{\alpha\rho}{}^\nu)^+g_{\beta\nu}^+=0
\ee
Here, with obvious meaning of the symbols, we denote: $g_{\beta\rho}^+=(g_{\beta\rho})^+$, $\overline{g}_{\beta\rho}=\overline{g_{\beta\rho}}$, etc. Consequently on $\Sigma$, from (\ref{conversejumps}$)_1$ we have:

\be
\ba{l}
\de_\alpha \overline{g}_{\beta\rho} +(1/2)\de_\alpha [g_{\beta\rho}]-\overline{\Gamma}_{\alpha\beta}{}^\nu \overline{g}_{\nu\rho}-\overline{\Gamma}_{\alpha\rho}{}^\nu
\overline{g}_{\nu\beta} + \cr
-(1/2)([\Gamma_{\alpha\beta}{}^\nu]\overline{g}_{\nu\rho}+\overline{\Gamma}_{\alpha\beta}{}^\nu[g_{\rho\nu}]+[\Gamma_{\alpha\rho}{}^\nu]\overline{g}_{\nu\beta}+\overline{\Gamma}_{\alpha\rho}{}^\nu[g_{\beta\nu}]) + \cr
-(1/4)([\Gamma_{\alpha\beta}{}^\nu][g_{\nu\rho}]+[\Gamma_{\alpha\rho}{}^\nu][g_{\beta\nu}])=0
\ea
\label{Dgplus}
\ee
Similarly, from the limit $f\longrightarrow 0^-$ and from (\ref{conversejumps}$)_2$ we also have on $\Sigma$:

\be
\ba{l}
\de_\alpha \overline{g}_{\beta\rho} -(1/2)\de_\alpha [g_{\beta\rho}]-\overline{\Gamma}_{\alpha\beta}{}^\nu \overline{g}_{\nu\rho}-\overline{\Gamma}_{\alpha\rho}{}^\nu
\overline{g}_{\nu\beta} + \cr
+(1/2)([\Gamma_{\alpha\beta}{}^\nu]\overline{g}_{\nu\rho}+\overline{\Gamma}_{\alpha\beta}{}^\nu[g_{\rho\nu}]+[\Gamma_{\alpha\rho}{}^\nu]\overline{g}_{\nu\beta}+\overline{\Gamma}_{\alpha\rho}{}^\nu[g_{\beta\nu}]) + \cr
-(1/4)([\Gamma_{\alpha\beta}{}^\nu][g_{\nu\rho}]+[\Gamma_{\alpha\rho}{}^\nu][g_{\beta\nu}])=0
\ea
\label{Dgminus}
\ee
From the sum of expressions (\ref{Dgplus}) and (\ref{Dgminus}) we thus have:

\be
\nabla_\alpha\overline{g}_{\beta\rho}
=(1/4)([\Gamma_\alpha{}_\beta{}^\nu][g_\nu{}_\rho]+[\Gamma_{\alpha\rho}{}^\nu][g_{\beta\nu}])
\label{Dgmean}
\ee
and, from difference: 

\be
\de_\alpha[g_{\beta\rho}]=[\Gamma_{\alpha\beta\rho}]+[\Gamma_{\alpha\rho\beta}]
\label{dgGG}
\ee
From (\ref{dgGG}), (\ref{product}), and from the definition of covariant derivative over $\Sigma$, we then have:

\be
\nabla_\alpha[g_{\beta\rho}]=[\Gamma_{\alpha\beta}{}^\nu]\overline{g}_{\nu\rho}+[\Gamma_{\alpha\rho}{}^\nu]\overline{g}_{\beta\nu}
\label{Dgjump}
\ee
As for the jump and the mean value of the Christoffel symbols we have, from (\ref{product}):
\be
\ba{lcl}
\overline{\Gamma}_{\alpha\beta}{}^\nu&=& (1/2)\bigl\{\overline{g}^{\nu\sigma}
(\de_\alpha \overline{g}_{\beta\sigma} +\de_\beta \overline{g}_{\sigma\alpha}-\de_\sigma \overline{g}_{\alpha\beta}
) + 
\cr && +
(1/4) [g^{\nu\sigma}](\de_\alpha [g_{\beta\sigma}] +\de_\beta [g_{\sigma\alpha}]
-\de_\sigma [g_{\alpha\beta}]
)\bigl\}
\ea
\ee 
and
\be
\ba{lcl}
[{\Gamma}_{\alpha\beta}{}^\nu]&=&(1/2)\bigl\{\overline{g}^{\nu\sigma}(\de_\alpha [g_{\beta\sigma}] +\de_\beta [g_{\sigma\alpha}]
-\de_\sigma [g_{\alpha\beta}]
)+
\cr && +
[g^{\nu\sigma}](\de_\alpha \overline{g}_{\beta\sigma} +\de_\beta \overline{g}_{\sigma\alpha}-\de_\sigma \overline{g}_{\alpha\beta}
)
\ea
\ee
or, from (\ref{Dgmean}) and (\ref{Dgjump}):

\be
\ba{lcl}
[\Gamma_{\alpha\beta}{}^\nu]\overline{g}_{\nu\rho}&=&(1/2)(\nabla_\alpha[g_{\beta\rho}]
+\nabla_\beta[g_{\rho\alpha}]-\nabla_\rho [g_{\alpha\beta}])\cr
[\Gamma_{\alpha\beta}{}^\nu][g_{\nu\rho}]&=&2(\nabla_\alpha \overline{g}_{\beta\rho}
+\nabla_\beta\overline{g}_{\rho\alpha}-\nabla_\rho \overline{g}_{\alpha\beta})
\ea
\ee

\section{Mean-value geometry on a hypersurface}

Let us consider a 4-vector $V^\alpha$, regularly discontinuous on $\Sigma$, the jump and the mean value of which will work as a
prototype of vectors with $\Sigma$ as support. We have, by definition:
\be
[\nabla_\beta\nabla_\alpha V^\sigma]=\nabla_\beta [\nabla_\alpha V^\sigma]
-[\Gamma_{\beta\alpha}{}^\nu]\overline{\nabla_\nu V^\sigma}+[\nabla_\beta\nu{}^\sigma]
\overline{\nabla_\alpha V^\nu}
\ee
where $[\nabla_\alpha V^\sigma]= \nabla_\alpha [V^\sigma]+[\Gamma_{\alpha\nu}{}^\sigma]\overline{V}^\nu$ and where, again by definition, we have:

\be
\overline{\nabla_\nu V^\sigma}=\frac{1}{2}\bigl\{
\de_\nu (V^+)^\sigma+(\Gamma^+)_{\nu\lambda}{}^\sigma (V^+)^\lambda +
\de_\nu (V^-)^\sigma+(\Gamma^-)_{\nu\lambda}{}^\sigma (V^-)^\lambda 
\bigr\}
\ee
Thus, from (\ref{conversejumps}) we have:

\be
\overline{\nabla_\nu V^\sigma}=\nabla_\nu \overline{V^\sigma}+({1}/{4})[\Gamma_{\nu\lambda}{}^\sigma][V^\lambda],
\ee
which, incidentally, is the same result we could get from the formal application of 
(\ref{product}), wich actually can be applied to covariant derivatives, provided one interpretes $\overline{\nabla}=\nabla$. 
We therefore have:

\be
\ba{lcl}
[\nabla_\alpha\nabla_\beta V^\sigma]&=& \nabla_\alpha\nabla_\beta [V^\sigma ]
+\nabla_\beta[\Gamma_{\alpha\nu}{}^\sigma]\overline{V}^\nu+[\Gamma_{\alpha\nu}{}^\sigma]
\nabla_\beta\overline{V}^\nu + \cr
&& -[\Gamma_{\beta\alpha}{}^\nu]\nabla_\nu\overline{V}^\sigma -({1}/{4})[\Gamma_{\beta\alpha}^{}\nu][\Gamma_{\nu\lambda}{}^\sigma] [V^\lambda] +\cr
&& 
+[\Gamma_{\beta\nu}{}^\sigma]\nabla_\alpha\overline{V}^\nu +
({1}/{4})[\Gamma_{\beta\nu}{}^\sigma][\Gamma_{\alpha\lambda}{}^\nu] [V^\lambda] 
\ea
\ee
and thus, by antisymmetrization:
\be
[\nabla_{[\beta}\nabla_{\alpha]}V^\sigma]= \nabla_{[\beta}\nabla_{\alpha]}[V^\sigma]
+ \nabla_{[\beta}[\Gamma_{\alpha]\nu{}^\sigma}]\overline{V}^\nu + \frac{1}{4}
[\Gamma_{\nu[\beta}{}^\sigma][\Gamma_{\alpha]\lambda}{}^\nu][V^\lambda]
\ee
Now, from the Ricci identity we have: $[\nabla_{[\beta}\nabla_{\alpha]}V^\sigma]=[R_{\alpha\beta\rho}{}^\sigma V^\rho]$
and then, by (\ref{product}):

\be
[\nabla_{[\beta}\nabla_{\alpha]}V^\sigma]=[R_{\alpha\beta\rho}{}^\sigma]\overline{V}^\rho+ \overline{R}_{\alpha\beta\rho}{}^\sigma [V^\rho],
\ee
and thus from a well known identity which follows from (\ref{product})
as a consequence our definition (\ref{covariantj}) for the covariant derivative on $\Sigma$, i.e. (see \cite{iogrg2002}):
\be
[R_{\alpha\beta\rho}{}^\sigma]=\nabla_\beta [\Gamma_{\alpha\rho}{}^\sigma]
-\nabla_\alpha [\Gamma_{\beta\rho}{}^\sigma]
\label{ideRG}
\ee
we have that the commutator of the covariant derivatives of the jump of a generic regularly discontinuous vector obeys the following Ricci-like formula:

\be
\nabla_{[\beta\alpha]}[V^\sigma] = \Bigl\{({1}/{2})\overline{R}_{\alpha\beta\rho}{}^\sigma
-({1}/{4})[\Gamma_{\nu[\beta}{}^\sigma][\Gamma_{\alpha]\rho}{}^\nu]\Bigl\}[V^\rho].
\ee
Not surprisingly, working in a similar way starting from $\overline{\nabla_\beta\nabla_\alpha V^\sigma}$ and antisymmetrizing, we find again:

\be
\nabla_{[\beta\alpha]}\overline{V}^\sigma = \Bigl\{({1}/{2})\overline{R}_{\alpha\beta\rho}{}^\sigma
-({1}/{4})[\Gamma_{\nu[\beta}{}^\sigma][\Gamma_{\alpha]\rho}{}^\nu]\Bigl\}\overline{V}^\rho;
\ee
in fact any given field with support on $\Sigma$ can be considered, by the help of suitable prolongations, as the jump (or as the mean value of) some regularly discontinuous field. Thus, for any vector $V$ with support on $\Sigma$, 
we can introduce the following mean-value geometry Ricci-like formula on $\Sigma$:

\be
(\nabla_{[\beta}\nabla_{\alpha]})V^\sigma = (\overline{R}_\Sigma)_{\alpha\beta\rho}{}^\sigma V^\rho;
\ee
where we have introduced the {\it mean-value geometry curvature} $(\overline{R}_\Sigma)$, defined by the following {\it mean-value geometry first Gauss-Codazzi identity}:

\be
(\overline{R}_\Sigma)_{\alpha\beta\rho}{}^\sigma= \overline{R}_{\alpha\beta\rho}{}^\sigma
-({1}/{4})\bigl([\Gamma_{\beta\nu}{}^\sigma][\Gamma_{\alpha\rho}{}^\nu]
-[\Gamma_{\alpha\nu}{}^\sigma][\Gamma_{\beta\rho}{}^\nu\bigr])
\label{mvg}
\ee
Notice that, for the sake of simplicity, we have introduced a slight abuse of notation, since
in \cite{iogrg2002} and \cite{iojgp2002} the same symbol $R_\Sigma$ instead denotes the inner curvature defined with the help of projections. Actually anything goes like in \cite{iogrg2002} section 4 with the Gauss-Codazzi framework, with the difference that here we don't have to make projections, which would involve product times a discontinuous tangent metric. Moreover here we don't even have to distinguish between the cases of $\Sigma$ timelike or lightlike. In other words our mean-value differential geometry on a hypersurface is a very simple, in conceptual terms, analogue of the Gauss-Codazzi apparatus. 

Thus, with the heuristic theory of \cite{iogrg2002} section 6 (see also \cite{israel66} for the timelike case) in mind as a prototype, we expect the jump of the Christoffel symbols to play the main role, in place of the secund fundamental form, in the definition of compatibility conditions for very weak solutions of the Einstein equations. Indeed, this happens, as it will be shown in the following.

\section{Complex mean-value formalism}

The metric being dicontinuous on $\Sigma$, we are missing the fundamental tool to rise and lower
indices, and to construct curvature in the traditional way. This is the reason why sometimes one is tempted to introduce some hybrid metric object on $\Sigma$ to replace the metric, even in the ($C^0$, piecewise $C^1$) case (see e.g. \cite{barra91}). It is reassuring to find out that the framework of the preceeding section can be confirmed by such a kind of approach.

It would be desiderable to simply replace $g$ with $\overline{g}$ on $\Sigma$, but it is easy to check that $\overline{g}$ has not the necessary algebraic requisites; in particular we have $\overline{g}_{\alpha\beta}\overline{g}^{\alpha\rho}\not = \delta_\beta{}^\rho$. Consider instead:

\be
\ba{ll}
\tilde{g}_{\alpha\beta}=\overline{g}_{\alpha\beta}+i(1/2)[g_{\alpha\beta}], &
\tilde{g}^{\alpha\beta}=\overline{g}^{\alpha\beta}-i(1/2)[g^{\alpha\beta}]
\ea
\label{gtild}
\ee 
where $i$ is the imaginary unit (i.e. we have $i^2=-1$). It is easy to check, with the help of (\ref{product}), that we have:

\be
\tilde{g}_{\alpha\beta}\tilde{g}^{\alpha\rho}=\delta_\alpha{}^\rho +i [g_{\alpha\beta}]\overline{g}^{\alpha\rho}
\ee
i.e., in particular: $\Re (\tilde{g}_{\alpha\beta}\tilde{g}^{\alpha\rho})=\delta_\beta{}^\rho$. 
For the sake of brevity in the following we will denote $A\approx B$ the relation $\Re(A)=\Re(B)$.
Thus the pair $\tilde{g}_{\alpha\beta}$ and $\tilde{g}^{\alpha\beta}$ is a good candidate replacement for the metric on $\Sigma$, for the purposes of rising and lowering indices. Now, similar to (\ref{gtild}) let us introduce:

\be
\ba{ll}
\tilde{\Gamma}_{\alpha\beta\nu}=\overline{\Gamma}_{\alpha\beta\nu}+i(1/2)[\Gamma_{\alpha\beta\nu}], & \tilde{\Gamma}_{\alpha\beta}{}^\sigma=\overline{\Gamma}_{\alpha\beta}{}^\sigma-i(1/2)[\Gamma_{\alpha\beta}{}^\sigma]
\ea
\ee
so that we have: $\tilde{\Gamma}_{\alpha\beta}{}^\sigma\approx\tilde{\Gamma}_{\alpha\beta\nu}\tilde{g}^{\sigma\nu}$
and conversely: $\tilde{\Gamma}_{\alpha\beta\nu}\approx \tilde{\Gamma}_{\alpha\beta}{}^\sigma\tilde{g}_{\nu\sigma}$. Let us now introduce the differential operator $\tilde{\nabla}$ on $\Sigma$, which makes use of $\tilde{\Gamma}$ in place of $\Gamma$.
As we could expect we have:

\be
\ba{ll}
\tilde{\nabla}_\rho\tilde{g}_{\alpha\beta}\approx 0, & \tilde{\nabla}_\rho\tilde{g}^{\alpha\beta}\approx 0
\ea
\ee
which is the replacement on $\Sigma$ for the covariant conservation of the metric tensor.

Now let us construct on $\Sigma$ the complex curvature tensor $\tilde{R}$ in the familiar way, but with $\tilde{\Gamma}$ in place of the ordinary Christoffel symbols (which are undefined on $\Sigma$):

\be
\tilde{R}_{\alpha\beta\rho}{}^\sigma = \de_\beta \tilde{\Gamma}_{\alpha\rho}{}^\sigma
-\de_\alpha \tilde{\Gamma}_{\beta\rho}{}^\sigma +\tilde{\Gamma}_{\beta\mu}{}^\sigma \tilde{\Gamma}_{\alpha\rho}{}^\mu -\tilde{\Gamma}_{\alpha\mu}{}^\sigma \tilde{\Gamma}_{\beta\rho}{}^\mu
\ee
We rather unespectedly find out that 

\be
\tilde{R}_{\alpha\beta\rho}{}^\sigma
= (\overline{R}_\Sigma)_{\alpha\beta\rho}{}^\sigma + i(1/2)[R_{\alpha\beta\rho}{}^\sigma]
\ee
i.e. in particular we have: $\tilde{R}_{\alpha\beta\rho}{}^\sigma\approx (\overline{R}_\Sigma)_{\alpha\beta\rho}{}^\sigma$, where
$\overline{R}_\Sigma$ is given by (\ref{mvg}). This is just another reason for identifying $R_\Sigma$ as the 
replacement for the curvature tensor of $\Sigma$, which is the first step of our path to the generalized compatibility conditions. 

\section{Generalized compatibility conditions}

Let us now consider limit $f\rightarrow 0^+$ of the curvature tensor of the subdomain $\Omega^+$;
by (\ref{conversejumps}) we have:

\be
(R_{\alpha\beta\rho}{}^\sigma)^+=\overline{R}_{\alpha\beta\rho}{}^\sigma +(1/2)[R_{\alpha\beta\rho}{}^\sigma]
\ee
and, by (\ref{ideRG}):

\be
(R_{\alpha\beta\rho}{}^\sigma)^+=
\overline{R}_{\alpha\beta\rho}{}^\sigma +\nabla_{[\beta}[\Gamma_{\alpha]\rho}{}^\sigma] 
\ee
We also have, by (\ref{mvg}):
\be
(R_{\alpha\beta\rho}{}^\sigma)^+=
(\overline{R}_\Sigma)_{\alpha\beta\rho}{}^\sigma
+\nabla_{[\beta}[\Gamma_{\alpha]\rho}{}^\sigma]+[\Gamma_{\nu[\beta}{}^\sigma][\Gamma_{\alpha]\rho}{}^\nu]
\ee      
We see that $\overline{R}$ and $\overline{R}_\Sigma$ only differ by terms proportional to $[\Gamma]$, and not involving derivatives of it.
Thus, in view of neglecting these tems, in the following we will consider $\overline{R}$ instead of $\overline{R}_\Sigma$; this simply avoids the introduction of the symbol $``\cong "$, with the meaning of equality but for terms not involving derivatives of $[\Gamma]$
(which here replaces the second fundamental form $K$)
as in \cite{iogrg2002} section 6.
Then for the Ricci tensor $R_{\beta\rho}=R_{\alpha\beta\rho}{}^\alpha$ we have:

\be
(R_{\beta\rho})^+= \overline{R}_{\beta\rho} + (1/2)\nabla_\mu \Bigl(\delta_\beta{}^\mu[\Gamma_\nu{}_\rho{}^\nu]-[\Gamma_{\beta\rho}{}^\mu]\Bigr)
\ee
and for the curvature scalar $R=R_\alpha{}^\alpha$:

\be
R^+=\overline{R}+(1/2)\nabla_\mu\Bigl([\Gamma_\nu{}^{\mu\nu}]-[\Gamma_\nu{}^{\nu\mu}]\Bigr)
\ee
Now, to construct the Einstein tensor $G^+$ we have to remember that, since the metric is also discontinuous:

\be
(g_{\alpha\beta})^+=\overline{g}_{\alpha\beta}+(1/2)[g_{\alpha\beta}]
\ee
so that we have:

\be
(G_{\beta\rho})^+=\overline{G}_{\beta\rho}+(1/2)\nabla_\mu \Bigl\{{\cal H}_{\beta\rho}{}^\mu
-(1/8)[g_{\beta\rho}]\Bigl([\Gamma_\nu{}^{\mu\nu}]-[\Gamma_\nu{}^\nu{}^\mu] \Bigr)\Bigr\}
\ee
where we have denoted, for the sake of brevity:

\be
{\cal H}_{\beta\rho}{}^\mu=
\Bigl\{ \delta_\beta{}^\mu[\Gamma_{\nu\rho}{}^\nu]-[\Gamma_{\beta\rho}{}^\mu]
-(1/2)\overline{g}_{\beta\rho}
\Bigl([\Gamma_\nu{}^{\mu\nu}]-[\Gamma_\nu{}^\nu{}^\mu]\Bigr)\Bigl\}
\label{defHgener}
\ee
Let us fix a coordinate chart and consider a generic (for the moment) regular prolongation for $G$, so that its mean value is defined in the whole $\Omega$. Now consider the Riemann 4-volume integral of $G^+$ over the domain $\Omega^+$; from the Green theorem we obtain (for the general definition of integral on a hypersurface see \cite{lich94} p. 6):  

\be
\int_{\Omega^+}G_{\beta\rho}=\int_{\Omega\backslash\Sigma}\overline{G}_{\beta\rho}
+(1/2)\int_{\Sigma}\ell^+_\mu {\cal H}_{\beta\rho}{}^\mu -(1/8)\int_{\Sigma}\ell^+_\mu [g_{\beta\rho}]
\Bigl([\Gamma_\nu{}^{\mu\nu}]-[\Gamma_\nu{}^\nu{}^\mu] \Bigr)
\ee
The analogous formula for $\Omega^-$ involves $-\ell^-$ as the outgoing normal vector and the
metric $g_{\alpha\beta}^-=\overline{g}_{\alpha\beta}-(1/2)[g_{\alpha\beta}]$, 
so we have:

\be
\int_{\Omega^-}G_{\beta\rho}=\int_{\Omega\backslash\Sigma}\overline{G}_{\beta\rho}
+(1/2)\int_{\Sigma}\ell^-_\mu {\cal H}_{\beta\rho}{}^\mu +(1/8)\int_{\Sigma}\ell^-_\mu [g_{\beta\rho}]
\Bigl([\Gamma_\nu{}^{\mu\nu}]-[\Gamma_\nu{}^\nu{}^\mu] \Bigr)
\ee
and consequently we have:
\be
\int_{\Omega}G_{\beta\rho}=\int_{\Omega\backslash\Sigma}\overline{G}_{\beta\rho}
+\int_{\Sigma}\overline{\ell}_\mu {\cal H}_{\beta\rho}{}^\mu 
\ee
Thus reasons similar to those of the heuristic theory (see \cite{israel66} and \cite{iogrg2002} section 6) lead to the reasonable hypothesis that $\overline{G}$ remain bounded in the neighbourhood of $\Sigma$, for any admissible prolongation,
so that from the volume integral of the Einstein equations, with the presence of an eventual source term concentrated on $\Sigma$:

\be
\int_\Omega G_{\alpha\beta} = -\chi \int_{\Omega\backslash \Sigma}T_{\alpha\beta}-\chi \int_\Sigma \breve{T}_\alpha{}_\beta
\ee
where $\chi$ denotes the gravitational constant, we conclude that 
\be
\int_\Sigma \overline{\ell}_\mu {\cal H}_{\beta\rho}{}^\mu =-\chi \int_\Sigma\breve{T}_{\beta\rho}
\ee
which is our heuristic reason for considering the following set of generalized compatibility conditions to
hold on $\Sigma$ as a replacement for the Einstein equations:

\be
\overline{\ell}_\mu {\cal H}_{\beta\rho}{}^\mu =-\chi \breve{T}_{\beta\rho}
\label{genecompa}
\ee
Here $\breve{T}$ represents the stress-energy content of the hypersurface. 

In the simpler case $\ell_\alpha\in C^0$, it is very easy to check that the object ${\ell}_\mu {\cal H}_{\beta\rho}{}^\mu$ is gauge-invariant in the sense of (\ref{gauge}), as we could hope.

Turning now to the comparison with the $C^0$ case,
we see from eq.s (71) and (85) of \cite{iogrg2002} that our generalized conditions (\ref{genecompa}) are formally identical to ordinary compatibility conditions [eq. (110) of the same paper], if expressed in terms of $[\Gamma]$ (which in the general case is a function of the jump of the metric $[g]$ as well as of its weak discontinuity $\de g$). Therefore it is clear that generalized compatibility conditions reduce to ordinary ones in case the metric is continuous, i.e. in case $[g_{\alpha\beta}]=0$.

In particular, let us suppose $g\in$($C^0$, piecewise $C^1$) and $f\in C^0(\Omega )$; let us moreover suppose $(\ell\cdot\ell)>0$, i.e. $\Sigma$ timelike. By definition of Christoffel symbols, and from (11) of \cite{iogrg2002}, we have:

\be
[\Gamma_{\beta\rho}{}^\sigma]=(\ell\cdot\ell)^{-1/2}(N_\beta\calG_\rho{}^\sigma+N_\rho\calG_\beta{}^\sigma -N^\sigma\calG_{\beta\rho})+(\ell\cdot\ell)^{1/2}N_\beta N_\rho Q^\sigma
\ee  
$Q$ is a vector which can be set to zero with a suitable gauge choice; it plays no role in (\ref{genecompa}), as one would expect, in fact we have:

\be
\ba{lcl}
\ell_\mu[\Gamma_{\beta\rho}{}^\mu]&=&-\calG_{\beta\rho}+(\ell\cdot\ell)N_{\beta\rho} (Q\cdot N) \cr
\ell_\beta [\Gamma_{\nu\rho}{}^\nu]&=& N_\beta N_\rho \calG_\nu{}^\nu +(\ell\cdot\ell)N_\beta N_\rho (Q\cdot N) \cr
\ell_\mu[\Gamma_\nu{}^{\mu\nu}]&=& \calG_\nu{}^\nu +(\ell\cdot\ell) (Q\cdot N)\cr
\ell_\mu[\Gamma_\nu{}^{\nu\mu}]&=& -\calG_\nu{}^\nu +(\ell\cdot\ell)(Q\cdot N)
\ea
\ee 
and, since $\overline{g}=g=h(N)+N\otimes N$, we have from (\ref{defHgener}):

\be
\ell_\mu {\cal H}_{\beta\rho}{}^\mu = \calG_{\beta\rho}-h(N)_{\beta\rho}\calG_\nu{}^\nu
\ee
i.e., according to (88) of \cite{iogrg2002}: 

\be
\ell_\mu {\cal H}_{\beta\rho}{}^\mu=\calH_{\beta\rho}
\ee
as expected.
Now let us instead suppose $(\ell\cdot\ell)=0$, i.e. $\Sigma$ lightlike. Let $u\in C^0(\Omega)$ be a given auxiliary reference frame. From eq.s (21)
and (16) of \cite{iogrg2002} we have:

\be
[\Gamma_{\rho\beta}{}^\sigma]= (u\cdot\ell)^{-1}(-L_\beta \calF(u)_\rho{}^\sigma-L_\rho \calF(u)_\beta{}^\sigma + L^\sigma \calF(u)_{\beta\rho})+(u\cdot\ell)^2 L_\beta L_\rho \hat{Q}^\sigma
\ee
and consequently, from (18) and (19) of the same paper:

\be
\ba{lcl}
\ell_\mu[\Gamma_{\beta\rho}{}^\mu]&=& L_\beta B(u,n)_\rho +L_\rho B(u,n)_\beta -(u\cdot\ell)^3
L_\beta L_\rho (\hat{Q}\cdot L)\cr
\ell_\beta [\Gamma_\nu{}_\rho{}^\nu]&=& L_\beta L_\rho \calG(u,n)_\nu{}^\nu -(u\cdot\ell)^3L_\beta L_\rho (\hat{Q}\cdot L)\cr
\ell_\mu[\Gamma_\nu{}^{\mu\nu}]&=& \ell_\mu [\Gamma_\nu{}^{\nu\mu}]=0
\ea
\ee
We therefore have:

\be
\ell_\mu\calH_{\beta\rho}{}^\mu = \calG(u,n)_\nu{}^\nu L_\beta L_\rho -L_\beta B(u,n)_\rho 
-L_\rho B(u,n)_\beta
\ee
i.e. again, according to (83) of \cite{iogrg2002}, we have: $\ell_\mu\calH_{\beta\rho}{}^\mu=\calH_{\beta\rho}$, as expected. 

Therefore the set (\ref{genecompa}) of compatibility conditions, together
with ordinary Einstein Equations to hold on each side of the discontinuity hypersurface, 
defines the class of generalized regularly discontinuous solutions of the Einstein equations. And in case $[g]=0$, i.e. for continuous metric, from such conditions we recover the ordinary
compatibility conditions for regularly discontinuous weak solutions.

However, in the generic case we have some differences, as we are going to show in the following.

\section{A class of spherical boundary layers}

Let us consider the match of two piecewise-$C^0$ regularly discontinuous spherical solutions of the vacuum Einstein equations, of the form
\be
ds^2=-a(r,t)dt^2+b(r,t)dr^2 +r^2d\Omega^2 
\label{sfer}
\ee
with $d\Omega^2=d\theta^2+\sin^2\theta d\varphi^2$, across a spherical 
admissible gravitational 
discontinuity hypersurface $\Sigma$ of equation $r=\rho(t)$, with $\rho(t)\in$ $C^1$. Therefore the form $\ell_\alpha= \delta_\alpha{}^r-\dot\rho\delta_\alpha{}^t$ is continuous (while $\ell^\beta=g^{\beta\alpha}\ell_\alpha$ in general is not). 
We suppose globally $C^0$ coordinates, the same form of the metric in both 
domains $\Omega^+$
and $\Omega^-$, and the identification $t^+=t^-$, $r^+=r^-$, $\theta^+=\theta^-$,
$\varphi^+=\varphi^-$ on $\Sigma$. Let$a,b>0$ and let $a,b\in$ piecewise-$C^0$ be regularly 
discontinuous on $\Sigma$ and with regularly discontinuous first derivatives. Let us denote by a dot the partial derivative with respect to $t$, and by a 
prime that with respect to $r$. Let moreover  condition $a-b\dot\rho>0$, i.e. $(\ell\cdot\ell)>0$, hold on both sides on $\Sigma$.

We have:
\be
[g_{\alpha\beta}] = -[a] \delta_\alpha{}^t \delta_\beta{}^t + [b]
\delta_\alpha{}^r \delta_\beta{}^r
\label{jgsfer}
\ee 
Now let us define the match as a generalized regularly discontinuous solution by (\ref{genecompa}), with $\breve T =0$, i.e. in the absence of stress-energy concentrated on $\Sigma$. In this case our compatibility conditions reduce to:

\be
{\ell}_\beta[\Gamma_{\mu\rho}{}^\mu]-{\ell}_\mu[\Gamma_{\beta\rho}{}^\mu]=0
\ee 
which, for a match of metrics of the kind (\ref{sfer}), are equivalent to the following system:

\be
\ba{rcl}
{\dot\rho}[\dot bb^{-1}]+[a'b^{-1}]&=&0\cr
{\dot\rho}[\dot ba^{-1}]+[a'a^{-1}]&=&0\cr
{\dot\rho}[a'a^{-1}]+[\dot a a^{-1}]&=&0\cr
{\dot\rho}[b'b^{-1}]+[\dot b b^{-1}]&=&0\cr
[b^{-1}]&=&0
\ea
\ee 
i.e. we have $[b]=0$ and consequently:  

\be
\ba{rcl}
{\dot\rho}[\dot b]+[a']&=&0\cr
{\dot\rho}[\dot ba^{-1}]+[a'a^{-1}]&=&0\cr
{\dot\rho}[a'a^{-1}]+[\dot a a^{-1}]&=&0\cr
{\dot\rho}[b']+[\dot b ]&=&0
\ea
\ee 
and from (\ref{product}):

\be
\ba{rcl}
{\dot\rho}[\dot b]+[a']&=&0\cr
({\dot\rho}\overline{\dot{b}}+\overline{a'})[a^{-1}]
&=&0\cr
({\dot\rho}\ \overline{a'}+\overline{\dot a})[a^{-1}]+
({\dot\rho}[a']+[\dot a])\overline{a^{-1}}&=&0\cr
{\dot\rho}[b']+[\dot b ]&=&0
\ea
\label{systemsfer}
\ee
Now if we had both ${\dot\rho}[\dot b]+[a']=0$ and ${\dot\rho}\overline{\dot{b}}+\overline{a'}=0$, by (\ref{conversejumps})
we would have 
${\dot\rho}\dot b+a'=0$ on both sides of the hypersurface.
We discard for the moment this singular situation, and from (\ref{systemsfer}$)_2$ we conclude that $[a^{-1}]=0$. 

Thus in this case our generalized compatibility conditions imply $[a]=[b]=0$, i.e. they force the match to be $C^0$, piecewise-$C^1$. 

In \cite{iogrg2002} we have already studied some examples of $C^0$, piecewise-$C^1$  matchs of metrics of the kind (\ref{sfer}) at a hypersurface of constant radius $r=r_b$, with $\ell_\alpha = \delta_\alpha{}^r$. Namely, we have considered: external Schwarzschild - internal Schwarzschild; external Schwarzschild - Tolman VI; external Schwarzschild - Tolman V. Such matchs obviously have $\ell_\alpha\in C^0$; moreover condition $\dot\rho\ \dot b+a'\not = 0$ reduce in this case to $a'\not=0$, which is obviously satisfied. In each case we have verified that 
condition $[a]=[b]=0$ imply $\de a=0$ (where $\de$ denotes first order discontinuity), which then define the match as a boundary layer  \cite{iogrg2002} (it actually also imply $\de b=0$, as one can verify). Such is a general result, since for a metric of the kind (\ref{sfer}) the completely temporal and radial components of the Einstein tensor
are independent from the second derivatives of the metric:

\be
\ba{lcl}
G_{tt}&=&-a(b'r+b^2-b)/r^2b^2\cr
G_{rr}&=&-({a'r-ab+a})/{ar^2}
\ea
\ee
so that the corresponding vacuum Einstein equations reduce to:

\be
\ba{l}
b'r+b^2-b=0, \cr
a'r-ab+a =0.
\ea
\label{bbaa}
\ee
Now, since in the match of (\ref{sfer}) vacuum solutions equations (\ref{bbaa}) are satisfied on each side of the interface $\Sigma$, their jump is in particular null, and from (\ref{product}) we have:

\be
\ba{l}
r_b[b']+(2\overline{b}-1)[b]=0 \cr
r_b[a']-\overline{a}[b]-(\overline{b}+1)[a]=0
\ea
\ee
from which it clearly follows that conditions $[a]=[b]=0$ imply $[a']=[b']=0$, i.e. $\de a=\de b=0$.

Summarizing, for the match of two piecewise-$C^0$ regularly discontinuous spherical solutions, in the above hypothesis, generalized compatibility conditions (\ref{genecompa}) imply $[a]=[b]=0$ i.e. they force 
the match to be $C^0$. On the other hand conditions $[a]=[b]=0$ imply that $\Sigma$ is a boundary layer. Therefore for such spherical matchs generalized compatibility conditions (\ref{genecompa}) are necessary and sufficient for the match to be a boundary layer.

\section{Generalized gravitational shock waves}

Let us consider the match of two plane wave metrics of the form
\be
ds^2=-2d\xi d\eta +F(\xi)^2dx^2 +G(\xi)^2dy^2   
\label{plane}
\ee
across a hypersurface $\Sigma$ of equation $\xi=0$.
Here $\xi$ and $\eta$ are the two null coordinates.
We suppose continuously matching coordinates and $F,G$
regularly 
discontinuous, together with their first and second derivatives. 
The gradient vector of $\Sigma$ is the continuous characteristic (on each side of $\Sigma$) vector 
$\ell_\alpha =\delta_\alpha{}^\xi$. 

Generalized compatibility conditions (\ref{genecompa}) in the case $\breve{T}=0$ (i.e. no stress-energy concentrated on the hypersurface) reduce to the following single scalar equation:

\be
[F^{-1}F'+G^{-1}G']=0
\label{compawave}
\ee
which characterize the generalized gravitational shock wave. Let us now study compatibility of (\ref{compawave}) with the Einstein Equations.
Einstein vacuum equations also reduce to a single scalar equation:

\be
F^{-1}F''+G^{-1}G''=0
\ee 
which we suppose to hold on each side of the hypersurface $\Sigma$; thus replacing $F^+$ and $G^+$ by their expressions in terms of $\overline{F}$, $[F]$, $\overline{G}$ and $[G]$ according to (\ref{conversejumps}) gives rise to the following couple scalar conditions:   

\begin{eqnarray}
(2\overline{F''}+[F''])(2 \overline{G}+[G])+(2\overline{G''}+[G''])(2\overline{F}+[F])=0\label{Ewave1}\\
(2\overline{F''}-[F''])(2\overline{G}-[G])+(2\overline{G''}-[G''])(2\overline{F}-[F])=0\label{Ewave2}
\end{eqnarray}
Equations (\ref{Ewave1})-(\ref{Ewave2}) are compatible with (\ref{compawave}), i.e. the three equations set can be solved algebraically with respect to $F$, $[G]$ and to any member of the pair ($\overline{F}$, $\overline{G}$), and the solution is not necessarily trivial. 

Finally let us notice that, if the additional condition $[F]=[G]=0$ holds, i.e. if the solution is $C^0$, condition (\ref{compawave}) reduces to
$F^{-1}[F']+G^{-1}[G']=0$
i.e.:

\be
F^{-1}\de F+G^{-1}\de G=0
\ee 
which is the analogous condition for the ordinary shock wave, according to \cite{iogrg2002} section 10.5.

\section{Slow generalized gravitational waves}

Let us start trying to match two vacuum solutions of the kind (\ref{plane}) across the timelike (on both sides) hypersurface $\Sigma$ of equation $\xi=\zeta$. Again we suppose continuously matching coordinates, $F,G$
regularly 
discontinuous together with their first and second derivatives, and
$\breve{T}=0$.
This times generalized compatibility conditions include (\ref{compawave}) and the following two additional scalar conditions:

\be
\ba{ll}
[FF']=0, & [GG']=0
\ea
\ee 
i.e., in terms of $\overline{F}$, $[F]$, $\overline{G}$ and $[G]$, according to (\ref{product}):

\begin{eqnarray}
\overline{F}[F']+[F]\overline{F'}=0 \label{compawavebis}\\
\overline{G}[G']+[G]\overline{G'}=0 \label{compawavetris}
\end{eqnarray}
It is easy to check that the system (\ref{compawavebis})-(\ref{compawavetris}) is not compatible with (\ref{Ewave1})-(\ref{Ewave2}), in the sense that the whole system does not admit non-trivial solutions for $\overline{F}$, $[F]$, $\overline{G}$ and $[G]$. 

On the other hand we have proved in section 6 that a wide class of generalized spherical matchs at a hypersurface of constant radius necessarily degenerate to a $C^0$ match. 

Other examples of degeneracy have not been included in the paper for the sake of brevity, but at least it seems to be a hard task to construct a non-trivial generalized match across a timelike (on each side) hypersurface, with no stress-energy content. Such difficulty is certainly not a proof that this is an impossible task, but it makes us wonder whether such a solution should necessarily degenerate to a boundary layer, just like it happens for ordinary $C^0$ solutions (see e.g. \cite{iogrg2002}). This would forbid the existence of generalized solutions which propagate at a speed slower than light. Such would be a desiderable prohibition under certain respect, since one could expect that gravitational interactions in vacuo must necessarily propagate at the speed of light also in a generalized theory. 

In general terms, since for generalized solutions the metric is discontinuous, a hypersurface can in principle have different signatures on the different sides; for this reason we cannot simply distinguish between the timelike and the lightlike case, as for usual $C^0$ solutions. We should rather distinguish between three cases: timelike-timelike, timelike-lightlike (or conversely) and lightlike-lightlike. 

In any case it is legitimate to expect that, at least in the timelike-timelike case, similar to the timelike case of ($C^0$, piecewise $C^1$) solutions, absence of stress-energy concentrated on $\Sigma$ should imply the solution to degenerate to a boundary layer \cite{iogrg2002}. 

Unfortunately for generalized solutions we still have no proof that absence of stress energy concentrated on $\Sigma$ does necessarily imply the degeneracy of the solution to a boundary layer.

Therefore, although the examples considered in this paper seem to suggest that such property could hold true also in the generalized case, for the moment such result is still a conjecture; we thus have to admit that the theory in principle allows propagation of generalized gravitational shock waves at lower speed than the speed of light. We would call such waves ``slow generalized gravitational shock waves". It would be reasonable to forbid this situation as unphysical, but for now this can only be done ad hoc, by means of a corresponding additional hypothesis. 

\section{Non-symmetric stress-energy}

Notice that   $\overline{\ell}_\mu\calH_{\beta\rho}{}^\mu$ is not necessarily symmetric; from identity:
\be
\Gamma_{\nu\alpha}{}^\nu=(1/2)g^{-1}\de_\alpha g
\ee 
where $g$ denotes the determinant of
the contravariant metric, we have:

\be
\overline{\ell}_\mu\calH_{[\beta\rho]}{}^\mu=(1/4)(\overline{\ell}_\beta [g^{-1}\de_\rho g]-\overline{\ell}_\rho
[g^{-1}\de_\beta g])
\label{mucaHg}
\ee
Thus the generalized scheme allows in principle the presence of non symmetric stress-energy
on the discontinuity hypersurface. We will display non-trivial examples of non-symmetry in the following section. Notice that the right hand side of (\ref{mucaHg}) is identically null in case $g\in C^0$ and $\ell_\alpha\in C^0$, since in this case we have $[g^{-1}\de_\beta g]=\ell_\beta g^{-1}\de g$.

A non-symmetric Einstein tensor is a feature of Einstein-Cartan theory of gravitation (see \cite{cartan22}, see also \cite{barrabeshoganbook} section 7.2), where it is due to the presence of torsion in the non-symmetric connection used to construct generalized curvature.  
Thus the generalized theory can be interpreted, at least to some extent, as introducing a torsion equivalent tool on the shell only, even if there are no geometrical objects in our theory which can be directly interpretated as torsion. However,
Einstein-Cartan theory also has a spin - angular momentum field equation in addition to the Einstein equations, which here is missing. 

In the literature, compatibility conditions for $C^0$ solutions of boundary layers \cite{arku}, and recently of shock waves and thin shells \cite{bressange}, have been studied also in the framework of Einstein-Cartan theory; actually this can lead to non-symmetric stress-energy on the shell. But in that theory this feature is inherited 
from the ambient spacetime, which is not here: non-symmetric stress-energy arises on the shell only, in consequence of the theory. This interesting feauture probably is worth investigating.

\section{Generalized thin shells}

Now let us consider a more general form of the spherical metric:
\be
ds^2=-a(r,t)dt^2+b(r,t)dr^2 +c(r,t)d\Omega^2 
\label{sfer2}
\ee
Let us consider a match of two spherical solutions of the Einstein equations across
a timelike (on each side) hypersurface of equation
$r=\rho(t)$. Again we suppose $\rho(t)\in$ $C^1$ and therefore $\ell_\alpha= \delta_\alpha{}^r-\dot\rho\delta_\alpha{}^t\in C^0$. Let the coordinates be $C^0$
globally, and let the metric have the same form (\ref{sfer2}) in both 
domains $\Omega^+$
and $\Omega^-$, with the identification $t^+=t^-$, $r^+=r^-$, $\theta^+=\theta^-$,
$\varphi^+=\varphi^-$ on $\Sigma$. 

Let moreover $a,b,c>0$ and let $a,b,c\in$ piecewise-$C^0$ be regularly 
discontinuous on $\Sigma$ and with regularly discontinuous first derivatives. Again we denote by a dot the partial derivative with respect to $t$, and by a 
prime that with respect to $r$.

In this case for the left hand side of the generalized compatibility conditions $\ell_\mu{\cal H}_{\beta\rho}{}^\mu$ we obtain:

\be
\ba{ll}
\ell_\mu{\cal H}_{\beta\rho}{}^\mu=& -\Bigl([a'b^{-1}/2]+\dot\rho[\dot b b^{-1}/2+\dot c c^{-1}]\Bigr)\delta_\beta{}^t\delta_\rho{}^t+ \\
&+ ([a'a^{-1}/2+c'c^{-1}]+\dot\rho[\dot ba^{-1}/2])\delta_\beta{}^r\delta_\rho{}^r+\\
&+([\dot a a^{-1}/2+\dot c c^{-1}]+\dot\rho[a'a^{-1}/2])\delta_\beta{}^r\delta_\rho{}^t +\\
&-([\dot b b^{-1}/2]+\dot\rho[b'b^{-1}/2+c'c^{-1}])\delta_\beta{}^t\delta_\rho{}^r +\\
&+([c'b^{-1}/2]+\dot\rho[\dot c a^{-1}/2])(\delta_\beta{}^\theta\delta_\rho{}^\theta+
\sin^2\theta \delta_\beta{}^\varphi\delta_\rho{}^\varphi)+ \\
&-\Bigl([b^{-1}(a'a^{-1}/2+c'c^{-1})]+\dot\rho[a^{-1}(\dot b b^{-1}/2+\dot c c^{-1})]\Bigr)
\overline{g}_{\beta\rho}
\ea
\label{gensfeshe}
\ee 
where, obviously:

\be
\overline{g}_{\beta\rho}=-\overline{a}\delta_\beta{}^t\delta_\rho{}^t+\overline{b}\delta_\beta{}^r\delta_\rho{}^r +\overline{c}(\delta_\beta{}^\theta\delta_\rho{}^\theta+
\sin^2\theta \delta_\beta{}^\varphi\delta_\rho{}^\varphi)
\ee
We now mean to interpret (\ref{gensfeshe}) as the matter-energy of a thin shell.
Let us first get back to the particular case $\dot\rho=0$ (static shell) 
and $\dot a=\dot b=\dot c=0$, to
make the interpretation simpler by eliminating the non-symmetric component; rearranging some terms we in fact obtain:

\be
\ba{ll}
\ell_\mu{\cal H}_{\beta\rho}{}^\mu=& (-[a'b^{-1}/2]+\overline{a}\overline{c}^{-1}[c'b^{-1}/2])\delta_\beta{}^t\delta_\rho{}^t+\\
&+([a'a^{-1}/2+c'c^{-1}]-\overline{b}\overline{c}^{-1}[c'b^{-1}/2])\delta_\beta{}^r\delta_\rho{}^r + \\
&+\Bigl(\overline{c}^{-1}[c'b^{-1}/2]-[b^{-1}(a'a^{-1}/2+c' c^{-1})]\Bigr)\overline{g}_{\beta\rho}
\ea
\label{lHsf}
\ee
This can be interpretated as a perfect isotropic magneto-fluid thin shell with 
infinite conductivity, i.e. we can solve the compatibility conditions by considering the following stress-energy as the right hand side:

\be
\breve{T}_{\alpha\beta}= (\rho_0 + p+\mu h^2)U_\alpha U_\beta  +(p+(1/2)\mu h^2)
g_{\alpha\beta} -\mu h_\alpha h_\beta
\label{pmf}
\ee 
where $\rho_0$ is the proper density, $h$ the magnetic field and $\mu$ the magnetic
permeability \cite{lich67,lich69,lich70,anile89,lich94}; here we define $h^2=h_\alpha h_\beta \overline{g}^{\alpha\beta}$.
In fact it suffices to define the following 4-velocity vector:
\be
U_\alpha=\ds\sqrt{\overline{a}-(1/4)[a^2]\overline{a}^{-1}}\delta_\alpha{}^t=(\overline{a^{-1}})^{1/2}\delta_\alpha{}^t
\ee
which by construction is unitary on $\Sigma$, in the following sense:
$U_\alpha U_\beta \overline{g}^{\alpha\beta}=-1$, and the following magneto-hydrodynamical variables:

\be
\ba{ll}
\rho_0=&\chi^{-1}\overline{a^{-1}}[a'b^{-1}]/2+\chi^{-1}\overline{c}^{-1}(\overline{b}\ \overline{b^{-1}}/2 -\overline{a}\overline{a^{-1}}+1)[c'b^{-1}]/2+\\
&-\chi^{-1}[b^{-1}](\overline{a'a^{-1}}/2+\overline{c'c^{-1}})-(3/2)\chi^{-1}\overline{b^{-1}}[a'a^{-1}/2+c'c^{-1}]\\
p=&\chi^{-1}\overline{c}^{-1}(\overline{b}\overline{b^{-1}}/2-1)[c'b^{-1}/2]+(1/2)\chi^{-1}\overline{b^{-1}}[a'a^{-1}/2+c'c^{-1}]+\\
&+\chi^{-1}[b^{-1}](\overline{a'a^{-1}}/2+\overline{c'c^{-1}})\\
h_\alpha=&\pm\sqrt{\overline{b^{-1}}\chi^{-1}([a'a^{-1}/2+c'c^{-1}]-\overline{b}\overline{c}^{-1}[c'b^{-1}/2])}\delta_\alpha{}^r
\ea
\label{sfershe}
\ee
to match (\ref{lHsf}) and (\ref{pmf}) via $\ell_\mu{\cal H}_{\beta\rho}{}^\mu=-\chi \breve{T}_{\beta\rho}$. If $[a]=[b]=[c]=0$ then the generalized shell (\ref{sfershe}) degenerates
to the $C^0$ magnetohydrodynamical shell considered in \cite{iogrg2002} section 10.1, in the particular case $\dot\rho=0$. 

The slightly more general case of $\dot a=\dot b =\dot c=0$, but $\dot\rho\not =0$, displays non-symmetric terms in (\ref{gensfeshe}); however it is not difficult to see that the perfect magnetofluid interpretation still holds, provided such
additional non-symmetric terms are interpreted, or neglected. In fact in this case we have:

\be
\ba{ll}
\ell_\mu{\cal H}_{\beta\rho}{}^\mu=& (-[a'b^{-1}/2]+\overline{a}\ \overline{c}^{-1}[c'b^{-1}/2])\delta_\beta{}^t\delta_\rho{}^t+\\
&+([a'a^{-1}/2+c'c^{-1}]-\overline{b}\overline{c}^{-1}[c'b^{-1}/2])\delta_\beta{}^r\delta_\rho{}^r + \\
&+(1/2)\dot\rho[a'a^{-1}/2-b'b^{-1}/2-c'c^{-1}](\delta_\beta{}^r\delta_\rho{}^t+\delta_\beta{}^t\delta_\rho{}^r)+ \\
&+(1/2)\dot\rho[a'a^{-1}/2+b'b^{-1}/2+c'c^{-1}](\delta_\beta{}^r\delta_\rho{}^t-\delta_\beta{}^t\delta_\rho{}^r)+ \\
&+\Bigl(\overline{c}^{-1}[c'b^{-1}/2]-[b^{-1}(a'a^{-1}/2+c' c^{-1})]\Bigr)\overline{g}_{\beta\rho}
\ea
\label{lHsfbis}
\ee
Now let us consider, for the sake of brevity, the following quantities:
\be
\alpha=\ds\frac{\frac{1}{4}\dot\rho^2[\frac{a'}{2a}-\frac{b'}{2b}-\frac{c'}{c}]^2\overline{b^{-1}}-(\frac{\overline{a}}{\overline{c}}[\frac{c'}{2b}]-[\frac{a'}{2b}])^2\overline{a^{-1}}}{\frac{\overline{a}}{\overline{c}}[\frac{c'}{2b}]-[\frac{a'}{2b}]}
\ee
\be
\beta=\biggl[\frac{a'}{2a}+\frac{c'}{c}\biggr]-\frac{\overline{b}}{\overline{c}}\biggl[\frac{c'}{2b}\biggr]
+\frac{1}{4}\dot\rho^2\frac{[\frac{a'}{2a}-\frac{b'}{2b}-\frac{c'}{c}]^2}{\frac{\overline{a}}{\overline{c}}[\frac{c'}{2b}]-[\frac{a'}{2b}]}
\ee
and let us suppose that inequality $\alpha<0$ holds, which is necessary for the physical interpretation. In fact in this case the following vector:

\be
U_\alpha = \frac{ 
(\frac{\overline{a}}{\overline{c}}[\frac{c'}{2b}]-[\frac{a'}{2b}])\delta_\alpha{}^t+\frac{1}{2}\dot\rho[\frac{a'}{2a}-\frac{b'}{2b}-\frac{c'}{c}]\delta_\alpha{}^r
}{
\sqrt{-\alpha (\frac{\overline{a}}{\overline{c}}[\frac{c'}{2b}]-[\frac{a'}{2b}])}}
\ee
is a unit timelike vector on $\Sigma$, in the sense that $U_\alpha U_\beta \overline{g}^{\alpha\beta}=-1$. Rearranging terms, (\ref{lHsfbis}) now reads:

\be
\ba{ll}
\ell_\mu{\cal H}_{\beta\rho}{}^\mu=&\alpha U_\beta U_\rho + \beta \delta_\beta{}^r\delta_\rho{}^r
+\frac{1}{2}\dot\rho[\frac{a'}{2a}+\frac{b'}{2b}+\frac{c'}{c}](\delta_\beta{}^r\delta_\rho{}^t-\delta_\beta{}^t\delta_\rho{}^r)+ \\
&+\Bigl(\overline{c}^{-1}[\frac{c'}{2b}]-[b^{-1}(\frac{a'}{2a}+\frac{c'}{c})]\Bigr)\overline{g}_{\beta\rho}
\ea
\ee
which can be matched via $\ell_\mu{\cal H}_{\beta\rho}{}^\mu=-\chi\breve{T}_{\beta\rho}$ with a stress-energy tensor of the following kind:

\be
\breve{T}_{\alpha\beta}= (\rho_0 + p+\mu h^2)U_\alpha U_\beta  +(p+(1/2)\mu h^2)
g_{\alpha\beta} -\mu h_\alpha h_\beta + A_{\alpha\beta}
\ee
where $A$ denotes the anti-symmetric term. We have:

\be
\ba{ll}
\rho_0=&-\chi^{-1}\alpha+\chi^{-1}\frac{1}{\overline{c}}[\frac{c'}{2b}]-\chi^{-1}[b^{-1}(\frac{a'}{2a}+\frac{c'}{c})]-\frac{1}{2\chi}\beta \overline{b^{-1}} \\
p=&-\chi^{-1}\frac{1}{\overline{c}}[\frac{c'}{2b}]+\chi^{-1}[b^{-1}(\frac{a'}{2a}+\frac{c'}{c})]-\frac{1}{2\chi}\beta \overline{b^{-1}} \\
\mu h^2 =&\chi^{-1}\beta \overline{b^{-1}}
\ea
\ee
while the anti-symmetric term $A$ reads:

\be
A_{\alpha\beta}=-\chi^{-1} \frac{1}{2}\dot\rho\biggl[\frac{a'}{2a}+\frac{b'}{2b}+\frac{c'}{c}\biggr](\delta_\beta{}^r\delta_\rho{}^t-\delta_\beta{}^t\delta_\rho{}^r)
\ee
The interpretation of such term is still missing; alternatively it could be neglegted by adding the further hypothesys:

\be
\biggl[\frac{a'}{2a}+\frac{b'}{2b}+\frac{c'}{c}\biggr]=0
\ee
which is equivalent to $[g^{-1}g']=0$, as we could expect from (\ref{mucaHg}).

\end{document}